\documentstyle[12pt,prd,aps,epsfig,preprint]{revtex}
\begin{document}
\title{{\bf Exact Eigenvalues and Eigenfunctions of the Hulth\'{e}n Potential in
the }${\rm PT-}${\bf Symmetry for Any Angular Momentum}}
\author{Sameer M. Ikhdair\thanks{%
sameer@neu.edu.tr} and \ Ramazan Sever\thanks{%
sever@metu.edu.tr}}
\address{$^{\ast }$Department of Electrical and Electronics Engineering, Near East\\
University, Nicosia, Cyprus.\\
$^{\dagger }$Department of Physics, Middle East Technical University,\\
Ankara, Turkey.}
\date{\today
}
\maketitle

\begin{abstract}
The Schr\"{o}dinger equation with the ${\rm PT-}$symmetric
Hulth\'{e}n potential is solved exactly by taking into account
effect of the centrifugal barrier for any $l$-state. Eigenfunctions
are obtained in terms of the Jacobi polynomials. The
Nikiforov-Uvarov method is used in the computations. Our numerical
results are in good agreement with the ones obtained before.\newline
Keywords: Energy Eigenvalues and Eigenfunctions; Hulth\'{e}n potential; $%
{\rm PT-}$symmetry; Nikiforov-Uvarov Method.\newline PACS Nos:
03.65.-w; 02.30.Gp; 03.65.Ge; 68.49.-h; 24.10.Ht; 03.65.Db;
12.39.Pn; 71.15.Dx; 02.30.Fn
\end{abstract}

\bigskip

% INITIALIZE - DONT CHANGE
%
%
%

\section{Introduction}

\noindent The Hulth\'{e}n potential [1] is a short-range potential in
physics. The potential form is

\begin{equation}
V^{(H)}(r)=-Ze^{2}\delta \frac{e^{-\delta r}}{1-e^{-\delta r}},
\end{equation}
where $Z$ is a constant and $\delta $ is the screening parameter. If the
potential is used for atoms, the $Z$ is identified with the atomic number.
This potential is a special case of the Eckart potential [2], which has been
widely used in several branches of physics and its bound states and
scattering properties have been investigated by a variety of techniques [3].

The radial Schr\"{o}dinger equation for the Hulth\'{e}n potential can be
solved analytically only for ${\rm s-}$states ($l=0)$ [1,4,5]$.$ For $l\neq
0,$ a number of methods have been used to find bound-state energies
numerically [3,6-16] and analytically [17,18]. In this context, we present
in this letter a Nikiforov-Uvarov method [19] within the frame of ${\rm PT-}$%
symmetric quantum mechanics [20-27] to solve analytically the Hulth\'{e}n
superpotential partner ($l+1)$th member for non-zero angular momentum states
given by Ref.[18]:

\begin{equation}
V_{\text{ }(l+1)}^{(H)}(r)=-Ze^{2}\delta \left[ 1-l(l+1)\frac{\hbar
^{2}\delta }{2mZe^{2}}\right] \frac{e^{-\delta r}}{1-e^{-\delta r}}+\frac{%
\hbar ^{2}\delta ^{2}}{2m}l(l+1)\frac{e^{-2\delta r}}{\left( 1-e^{-\delta
r}\right) ^{2}},\text{ }l=0,1,2,...
\end{equation}
which is for ${\rm s-}$states is not shape invariant [28] and leads to the
usual Hulth\'{e}n potential (1). On the other hand, Eq.(2) can be rearranged
as

\begin{equation}
V_{\text{ }eff}^{(H)}(r)=V_{(l+1)}(r)=-Ze^{2}\delta \frac{e^{-\delta r}}{%
1-e^{-\delta r}}+\frac{l(l+1)\hbar ^{2}\delta ^{2}}{2m}\frac{e^{-\delta r}}{%
\left( 1-e^{-\delta r}\right) ^{2}},
\end{equation}
which is well-known as the approximate Hulth\'{e}n effective potential
introduced by Greene and Aldrich [29].\footnote{%
In Ref.[30] the Hulth\'{e}n effective potential is taken as $V_{\text{ }%
eff}^{(H)}(r)=-\delta \frac{e^{-\delta r}}{1-e^{-\delta r}}+\frac{l(l+1)}{2}%
\left( \delta \frac{e^{-\delta r}}{1-e^{-\delta r}}\right) ^{2}$ in atomic
units.} It is of much concern to see that for small values of $\delta ,$
Eq.(3) becomes the effective Coulomb potential given by

\begin{equation}
V_{\text{ }eff}^{(H)}(r,\delta \approx 0)\rightarrow V_{\text{ }eff}^{(C)}=-%
\frac{Ze^{2}}{r}+\frac{l(l+1)\hbar ^{2}}{2mr^{2}}.
\end{equation}
In the case of Coulomb potential, the Hamiltonian corresponds to the
addition of an appropriate barrier potential and the so-called degenarcy is
recovered as a natural consequence [18]. At small values of $r$, the
Hulth\'{e}n potential, behaves like a Coulomb potential whereas for large
values of $r$ it decreases exponentially so its capacity for bound state is
smaller than the Coulomb potential. The coulomb problem is analytically
solvable for all energies and angular momenta. Because of the similarity and
points of contrast mentioned above between Coulomb and Hulth\'{e}n
potentials, it may be of much interest to use the Hulth\'{e}n superpotential
partner, Eq.(2), to generate their eigenvalues and eigenfunctions in the
framework of the ${\rm PT-}$symmetric quantum mechanics by employing
Nikiforov-Uvarov (${\rm NU}$) method. The outline of the paper is as
follows: In section II, we solve the Schr\"{o}dinger equation ${\rm SE}$
with Hulth\'{e}n superpotential for its energy eigenvalues and
eigenfunctions. In section III, we consider the $l=0$ and $l\neq 0$ cases
and compare with the other works and methods. Finally, section IV is for our
conclusions.

\section{Polynomial Solution of the Hulth\'{e}n potential}

\noindent The NU method provides us an exact solution of non-relativistic SE
for certain kind of potentials [19]. The method is based upon the solutions
of general second order linear differential equation with special orthogonal
functions [31]. For a given real or complex potentials, the ${\rm SE}$ in
one dimension, which is a ${\rm PT-}$symmetric equation, is reduced to a
generalized equation of hypergeometric type with an appropriate $s=s(x)$
coordinate transformation. Thus, it takes the form:

\begin{equation}
\psi ^{\prime \prime }(s)+\frac{\widetilde{\tau }(s)}{\sigma (s)}\psi
^{\prime }(s)+\frac{\widetilde{\sigma }(s)}{\sigma ^{2}(s)}\psi (s)=0,
\end{equation}
where $\sigma (s)$ and $\widetilde{\sigma }(s)$ are polynomials, at most of
second-degree, and $\widetilde{\tau }(s)$ is of a first-degree polynomial.
To find a particular solution for ${\rm SE}$ by separation of variables, we
use the transformation given by

\begin{equation}
\psi (s)=\phi (s)y(s).
\end{equation}
This reduces ${\rm SE}$ into an equation of hypergeometric type:

\begin{equation}
\sigma (s)y^{\prime \prime }(s)+\tau (s)y^{\prime }(s)+\lambda y(s)=0,
\end{equation}
where $\phi (s)$ is found to \i satisfy the condition $\phi ^{\prime
}(s)/\phi (s)=\pi (s)/\sigma (s).$ Further, $y(s)$ is the hypergeometric
type function whose polynomial solutions are given by Rodrigues relation

\begin{equation}
y_{n}(s)=\frac{B_{n}}{\rho (s)}\frac{d^{n}}{ds^{n}}\left[ \sigma ^{n}(s)\rho
(s)\right] ,
\end{equation}
where $B_{n}$ is a normalizing constant and the weight function $\rho (s)$
must satisfy the condition [19]

\begin{equation}
\left( \sigma (s)\rho (s)\right) ^{\prime }=\tau (s)\rho (s).
\end{equation}
The function $\pi (s)$ and the parameter $\lambda $ required for this method
are defined by
\begin{equation}
\pi (s)=\frac{\sigma ^{\prime }(s)-\widetilde{\tau }(s)}{2}\pm \sqrt{\left(
\frac{\sigma ^{\prime }(s)-\widetilde{\tau }(s)}{2}\right) ^{2}-\widetilde{%
\sigma }(s)+k\sigma (s)},
\end{equation}
and

\begin{equation}
\lambda =k+\pi ^{\prime }(s).
\end{equation}
Here, $\pi (s)$ is a polynomial with the parameter $s$ and the determination
of $k$ is the essential point in the calculation of $\pi (s).$ Thus, for the
determination of $k,$ the discriminant under the square root is being set
equal to zero and the resulting second-order polynomial has to be solved for
its roots $k_{1,2}$. Hence, a new eigenvalue equation for the SE becomes

\begin{equation}
\lambda _{n}+n\tau ^{\prime }(s)+\frac{n\left( n-1\right) }{2}\sigma
^{\prime \prime }(s)=0,\text{ \ \ \ \ \ \ }\left( n=0,1,2,...\right)
\end{equation}
where

\begin{equation}
\tau (s)=\widetilde{\tau }(s)+2\pi (s),
\end{equation}
and it will have a negative derivative.

Now, we follow Ref.[32], by rewritting Eq.(2) in a quite simple form as

\begin{equation}
V(r)=-V_{1}\frac{e^{-\delta r}}{1-e^{-\delta r}}+V_{2}\left( \frac{%
e^{-\delta r}}{1-e^{-\delta r}}\right) ^{2},\text{ \ }
\end{equation}
with

\begin{equation}
V_{1}=Ze^{2}\delta \left[ 1-l(l+1)\frac{\hbar ^{2}\delta }{2mZe^{2}}\right]
\text{ and \ }V_{2}=\frac{\hbar ^{2}\delta ^{2}}{2m}l(l+1).\text{\ }
\end{equation}
Therefore, using the separation of variables

\begin{equation}
\psi ({\bf r})=\frac{1}{r}R(r)Y(\theta ,\phi ),
\end{equation}
we may write the radial part of ${\rm SE}$ for all angular momentum states as

\begin{equation}
-\frac{\hbar ^{2}}{2m}\frac{d^{2}R(r)}{dr^{2}}+\left( V(r)+\frac{l(l+1)\hbar
^{2}}{2mr^{2}}\right) R(r)=ER(r).
\end{equation}
On the other hand, the one-dimensional counterpart of Eq.(17) can be
written, in a ${\rm PT-}$ symmetric form, as

\begin{equation}
R^{\prime \prime }(x)+\frac{2m}{\hbar ^{2}}\left[ E+\frac{V_{1}e^{-\delta x}%
}{1-e^{-\delta x}}-\frac{V_{2}e^{-2\delta x}}{\left( 1-e^{-\delta x}\right)
^{2}}\right] R(x)=0,
\end{equation}
and with the assi\c{s}gnment $s=e^{-\delta x},$ then it becomes

\begin{equation}
\frac{d^{2}R(s)}{ds^{2}}+\frac{1}{s}\frac{dR(s)}{ds}+\frac{2m}{\hbar
^{2}\delta ^{2}s^{2}}\left[ E+\frac{V_{1}s}{1-s}-\frac{V_{2}s^{2}}{\left(
1-s\right) ^{2}}\right] R(s)=0,
\end{equation}
and also introducing the given dimensionless parameters

\begin{equation}
\epsilon =-\frac{2mE}{\hbar ^{2}\delta ^{2}}>0\text{ \ \ (}E<0),\text{ \ \ }%
\beta =\frac{2mV_{1}}{\hbar ^{2}\delta ^{2}}\text{ \ \ (}\beta >0),\text{ \ }%
\gamma =\frac{2mV_{2}}{\hbar ^{2}\delta ^{2}}\text{ \ (}\gamma >0),
\end{equation}
finally it leads into the following simple hypergeometric form given by

\begin{equation}
\frac{d^{2}R(s)}{ds^{2}}+\frac{(1-s)}{s(1-s)}\frac{dR(s)}{ds}+\frac{1}{\left[
s\left( 1-s\right) \right] ^{2}}\times \left[ -\left( \epsilon +\beta
+\gamma \right) s^{2}+\left( 2\epsilon +\beta \right) s-\epsilon \right]
R(s)=0.
\end{equation}
Hence, comparing the last equation with the generalized hypergeometric type,
Eq.(5), we obtain the associated polynomials as

\begin{equation}
\widetilde{\tau }(s)=1-s,\text{ \ \ \ }\sigma (s)=s(1-s),\text{ \ \ }%
\widetilde{\sigma }(s)=-\left( \epsilon +\beta +\gamma \right) s^{2}+\left(
2\epsilon +\beta \right) s-\epsilon .
\end{equation}
When these polynomials are substituted into Eq.(10), with $\sigma ^{\prime
}(s)=1-2s,$ we obtain

\begin{equation}
\pi (s)=-\frac{s}{2}\pm \frac{1}{2}\sqrt{\left( 1+4\epsilon +4\beta +4\gamma
+4k\right) s^{2}-4\left( \beta +2\epsilon +k\right) s+4\epsilon }.
\end{equation}
Further, the discriminant of the upper expression under the square root has
to be set equal to zero. Therefore, it becomes

\begin{equation}
\Delta =\left[ 4\left( \beta +2\epsilon +k\right) \right] ^{2}-4\times
4\epsilon \left( 1+4\epsilon +4\beta +4\gamma +4k\right) =0.
\end{equation}
Solving Eq.(24) for the constant $k,$ we get the double roots as $%
k_{+,-}=-\beta \pm \sqrt{\epsilon \left( 1+4\gamma \right) },$ and
substituting these values for each $k$ into Eq.(23), we obtain

\begin{equation}
\pi (s)=-\frac{s}{2}\pm \frac{1}{2}\left\{
\begin{array}{c}
\left[ \left( 2\sqrt{\epsilon }-\sqrt{1+4\gamma }\right) s-2\sqrt{\epsilon }%
\right] ;\text{ \ \ \ for \ \ }k_{+}=-\beta +\sqrt{\epsilon \left( 1+4\gamma
\right) }, \\
\left[ \left( 2\sqrt{\epsilon }+\sqrt{1+4\gamma }\right) s-2\sqrt{\epsilon }%
\right] ;\text{ \ \ for \ \ }k_{-}=-\beta -\sqrt{\epsilon \left( 1+4\gamma
\right) }.
\end{array}
\right.
\end{equation}
Making the following choice for the polynomial $\pi (s)$ as

\begin{equation}
\pi (s)=-\frac{s}{2}-\frac{1}{2}\left[ \left( 2\sqrt{\epsilon }+\sqrt{%
1+4\gamma }\right) s-2\sqrt{\epsilon }\right] ,
\end{equation}
gives the function:

\begin{equation}
\tau (s)=1-2s-\left[ \left( 2\sqrt{\epsilon }+\sqrt{1+4\gamma }\right) s-2%
\sqrt{\epsilon }\right] ,
\end{equation}
which has a \ negative derivative of the form $\tau (s)=-\left( 2+2\sqrt{%
\epsilon }+\sqrt{1+4\gamma }\right) .$ Thus, from Eq.(11) and Eq.(12), we
find

\begin{equation}
\lambda =-\beta -\frac{1}{2}\left( 1+2\sqrt{\epsilon }\right) \left( 1+\sqrt{%
1+4\gamma }\right) ,
\end{equation}
and

\begin{equation}
\lambda _{n}=-\left( 2+2\sqrt{\epsilon }+\sqrt{1+4\gamma }\right) n-n(n-1).
\end{equation}
After setting $\lambda _{n}=\lambda $ and solving for $\epsilon ,$ we find:

\begin{equation}
\epsilon _{n}=\left[ \frac{1+2n}{2}-\frac{\left( n(n+1)+\beta \right) }{1+2n+%
\sqrt{1+4\gamma }}\right] ^{2}.
\end{equation}
which is exactly Eq.(26) in Ref.[32] for the deformed Woods-Saxon potential
if one lets $q=-1$ and $a=1/\delta .$ Therefore, substituting the values of $%
\epsilon ,$ $\beta $ and $\gamma $ into Eq.(30), one can immediately
determine the Hulth\'{e}n's exact energy eigenvalues $E_{n,l+1}$ as

\begin{equation}
E_{n,l+1}^{(H)}=-\frac{\hbar ^{2}\delta ^{2}}{2m}\left[ \frac{1+2n}{2}-\frac{%
\left( n(n+1)+\frac{2mV_{1}}{\hbar ^{2}\delta ^{2}}\right) }{1+2n+\sqrt{1+%
\frac{8mV_{2}}{\hbar ^{2}\delta ^{2}}}}\right] ^{2},\text{ \ \ }0\leq
n<\infty .\text{ \ }
\end{equation}
Therefore, substituting, Eq.(15) into Eq.(31), one gets

\begin{equation}
E_{n,l+1}^{(H)}=-\frac{\hbar ^{2}}{2m}\left[ \frac{\left( me^{2}Z/\hbar
^{2}\right) }{n+l+1}-\frac{\left( n+l+1\right) }{2}\delta \right] ^{2},\text{
\ \ }0\leq n<\infty .\text{ \ }
\end{equation}
for $l+1$ Hulth\'{e}n superpotential. Following Ref.[32], in atomic units ($%
\hbar =m=c=e=1)$ and for $Z=1,$ Eq.(32) turns out to be

\begin{equation}
E_{n,l+1}^{(H)}=-\frac{1}{2}\left[ \frac{1}{n+l+1}-\frac{(n+l+1)}{2}\delta %
\right] ^{2},\text{ \ \ }0\leq n<\infty ,\text{ \ \ }l=0,1,2,...
\end{equation}
or

\begin{equation}
E_{\overline{n},l}^{(H)}=-\frac{1}{2}\left[ \frac{1}{\overline{n}+l}-\frac{%
\overline{n}+l}{2}\delta \right] ^{2},\text{ \ \ }\overline{n}=n+1,\text{ \
\ }l=0,1,2,...
\end{equation}
which is exactly the same result obtained by other works (cf. e.g.,
Ref.[33], Eq.(78)) if $l$ is set equal to zero. The above equation indicates
that we deal with a family of Hulth\'{e}n potentials.\footnote{%
The critical screening $\delta _{c},$ at which $E_{n}=0,$ is defined, in
atomic units, by $\delta _{c}=2/(n+l+1)^{2}.$} Equation (32) agrees with
Eq.(5) in Ref.[18] for $l\neq 0$ case, and Eq.(11) in Ref.[30] for $l=0$
case. Of course, it is clear that by imposing appropriate changes in the
parameters $\delta $ and $V_{1},$ the index $n$ describes the quantization
for the bound energy states. In addition, if the parameter $V_{2}$ in
Eq.(31) is adjusted to zero, solution reduces to the form obtained for the
standard Hulth\'{e}n potential without a barrier term (cf. e.g., Eqs.(14)
and (15) with $l=0$ case).

Let us now find the corresponding wavefunctions. Applying the ${\rm NU}$
method, the polynomial solutions of the hypergeometric function $y(s)$
depends on the determination of weight function $\rho (s)$ which is found to
be

\begin{equation}
\rho (s)=(1-s)^{\eta -1}s^{2\sqrt{\epsilon }};\text{ \ \ \ \ }\eta =1+\sqrt{%
1+4\gamma }.
\end{equation}
Substituting into the Rodrigues relation given in Eq.(8), the eigenfunctions
are obtained in the following form

\begin{equation}
y_{n,q}(s)=C_{n}(1-s)^{-(\eta -1)}s^{-2\sqrt{\epsilon }}\frac{d^{n}}{ds^{n}}%
\left[ \left( 1-s\right) ^{n+\eta -1}s^{n+2\sqrt{\epsilon }}\right] ,
\end{equation}
where $C_{n}$ stands for the normalization constant and its value is $1/n!.$
The polynomial solutions of $\ y_{n}(s)$ are expressed in terms of Jacobi
Polynomials, which is one of the classical orthogonal polynomials, with
weight function $(1-s)^{\eta -1}s^{2\sqrt{\epsilon }}$ in the closed
interval $\left[ 0,1\right] ,$ yielding $A_{n}P_{n}^{(2\sqrt{\epsilon },\eta
-1)}(1-2s)$ [31]. Finally, the other part of the wave function in Eq.(6) is
found to be

\begin{equation}
\phi (s)=(1-s)^{\mu }s^{\sqrt{\epsilon }},\text{ \ \ }\mu =\eta /2.
\end{equation}
Combining the Jacobi polynomials and $\phi (s)$ in Eq.(36), the ${\rm s-}$%
wave functions ($l=0)$ could be determined as

\begin{equation}
R_{n}(s)=D_{n}s^{\sqrt{\epsilon }}(1-s)^{\mu }P_{n}^{(2\sqrt{\epsilon },\eta
-1)}(1-2s),
\end{equation}
with $s=e^{-\delta x}$ and $D_{n}$ is a new normalization constant.\ \ \ \ \
\ \ \ \ \ \ \ \ \ \ \ \ \ \ \ \ \ \ \ \ \ \ \ \ \ \ \ \ \ \ \ \ \ \ \ \ \ \
\ \ \ \ \ \ \ \ \

\section{Conclusions}

The exact solutions of the radial ${\rm SE}$ for the Hulth\'{e}n potential
with the angular momentum $l=0$ and $l\neq 0$ are found by using ${\rm NU}$
method. Eigenvalues and eigenfunctions obtained from the real form of the
potential are computed. Therefore, the wave functions are physical and
energy eigenvalues are in good agreement with the results obtained by the
other methods. In this regard, Figure 1 shows the variation of the
Hulth\'{e}n potential with $r$ for $S-$, $P-$, and $D-$states. Figure 2
plots the variation of the Hulth\'{e}n potential with $r$ for the $S-$state
with various screening parameters $\delta =0.002,0.01$ and $0.1$. Figures 3
and 4 show the variation of the energy eigenvalues with respect to the
quantum number $n$ for $S-$ and $P-$states with the a chosen values of
screening parameter $\delta =0.002,0.05$ and $0.2,$ respectively. On the
other hand, Table 1 shows the bound energy eigenvalues of the Hulth\'{e}n
potential as a function of $\delta $ for various quantum numbers of $S$%
-state. These results are compared with other works [11,33]. Further, the
bound energy eigenvalues as a function of $\delta $ for the states $2p$ ($%
n=0,l=1)$ and $3d\ (n=1,l=1)$ [30] are given in Table 2$.$
Comparison of our results with numerical data of Refs.[3,30] is also
given. However, since the form of the potential used in our work has
a different form than the potential form used by Ref.[30] in Eq.
(14), we have to add a perturbation term, $\Delta E,$ to our
calculations in order to substitute the small differences in [18].
Better results have been obtained for the $2p$ state for small
values of $\delta $ since the effective potential (3) becomes closer
to the original Hulth\'{e}n potential (1) and for small $l$ the
contribution of this angular momentum term in potential is also
small. Therefore, if all the parameters of potential remain purely
real, it is clear that all bound energies $E_{n}$ with $n\geq 0$
represent a negative energy spectrum [32]. We also point out that
the exact results obtained for the Hulth\'{e}n potential may have
some interesting applications in the study of different quantum
mechanical systems and atomic physics.\newline

ACKNOWLEDGMENTS\\
 This research was partially supported by the
Scientific and Technical Research Council of Turkey. The authors
thanks Ibrahim AbuAwwad for his assistance in drawing the Figures.
He acknowledges his wife, Oyoun, and also his son, Musbah, for their
love, encouragement and assistance. Their encouragement provided the
necessary motivation to complete this work.\newpage
\begin{figure}[tbp]
\caption{Variation of the Hulth\'{e}n potential as a function $r.$
The
curves are shown for screening parameter $\protect\delta =0.2$ for the $%
S-,P-,$ and $D-$states.} \label{Figure1}
\end{figure}

%\end{document}
\bigskip

\begin{figure}[tbp]
\caption{Variation of the Hulth\'{e}n potential as a function $r.$
The curves are shown for $S-$state with various values of the
screening parameter $\protect\delta =0.002,0.01$ and $0.1.$}
\label{Figure 2}
\end{figure}

%\end{document}

\begin{figure}[tbp]
\caption{The variation of the energy eigenvalues with respect to the
quantum number $n.$ The curves shown are for $S-$state with various
values of the screening parameters $\protect\delta =0.002,0.05$ and
$0.2.$} \label{Figure 3}
\end{figure}

%\end{document}

\bigskip

\begin{figure}[tbp]
\caption{The variation of the energy eigenvalues with respect to the
quantum number $n.$ The curves shown are for $P-$state with the same
screening parameters as in Figure 3.} \label{Figure 4}
\end{figure}

\bigskip

\bigskip \baselineskip= 2\baselineskip% double space the text

\bigskip

%\end{document}
\bigskip

\begin{table}[tbp]
\caption{The $S$-states energy eigenvalues of the Hulth\'{e}n potential for
several values of screening parameter $\protect\delta .$}
\label{Table 1}
\begin{tabular}{lllll}
$n$ & $-E_{n}\text{[11]}$ & $-\overline{E}_{n}\text{[33]}$ & $-E_{exact}$ & $%
\text{Our work}$ \\
\tableline &  & $\delta =0.002$ &  &  \\
1 & 0.4990005 & 0.4990005 & 0.4990005 & 0.4990005 \\
2 & 0.1240020 & 0.1240020 & 0.1240020 & 0.1240020 \\
3 & 0.0545601 & 0.0545601 & 0.0545601 & 0.0545601 \\
4 & 0.0302580 & 0.0302580 & 0.0302580 & 0.0302580 \\
5 &  & 0.0012500 &  & 0.0190125 \\
&  & $\delta =0.01$ &  &  \\
1 & 0.4950125 & 0.4950125 & 0.4950125 & 0.4950125 \\
2 & 0.1200500 & 0.1200500 & 0.1200500 & 0.1200500 \\
3 & 0.0506681 & 0.0506681 & 0.0506681 & 0.0506681 \\
4 & 0.0264501 & 0.0264500 & 0.0264500 & 0.0264500 \\
5 & 0.0153128 & 0.0153125 & 0.0153125 & 0.0153125 \\
&  & $\delta =0.05$ &  &  \\
1 & 0.4753125 & 0.4753125 & 0.4753125 & 0.4753125 \\
2 & 0.1012503 & 0.1012500 & 0.1012500 & 0.1012500 \\
3 & 0.0333746 & 0.0333681 & 0.0333681 & 0.0333681 \\
4 & 0.0113035 & 0.0112500 & 0.0112500 & 0.0112500 \\
5 &  & 0.0028125 &  & 0.0028125 \\
&  & $\delta =0.2$ &  &  \\
1 & 0.4049962 & 0.4050000 & 0.4050000 & 0.4050000 \\
2 & 0.0450856 & 0.0450000 & 0.0450000 & 0.0450000 \\
3 &  & 0.0005556 &  & 0.0005556 \\
4 &  & 0.0112500 &  & 0.0112500
\end{tabular}
\end{table}

\widetext

\bigskip

\begin{table}[tbp]
\caption{Energy eigenvalues as a function of the screening parameter $%
\protect\delta $ for the states $2p$ and $3d.$}
\label{Table 2}
\begin{tabular}{lllll}
State & $\delta $ & $-E_{n,l}$ [3]\tablenote{Variational method.} & $%
-E_{n,l} $ [30]\tablenote{Numerical integration.} & Our work\tablenote{The
small difference in results is because the potential form used by Ref.[30]
for $l\ne 0$ part is different than our form.} \\
\tableline$2p$ & 0.025 & 0.112760 & 0.1127605 & 0.1128125 \\
& 0.050 & 0.101042 & 0.1010425 & 0.1012500 \\
& 0.075 & 0.089845 & 0.0898478 & 0.0903125 \\
& 0.100 & 0.079170 & 0.0791794 & 0.0800000 \\
& 0.150 & 0.059495 & 0.0594415 & 0.0612500 \\
& 0.200 & 0.041792 & 0.0418860 & 0.0450000 \\
$3d$ & 0.025 & 0.043601 & 0.0437069 & 0.0437590 \\
& 0.050 & 0.032748 & 0.331645 & 0.0333681 \\
& 0.075 & 0.023010 & 0.0239397 & 0.0243837 \\
& 0.100 & 0.014433 & 0.0160537 & 0.0168056
\end{tabular}
\end{table}

\newpage

\begin{figure}
\epsfig{file=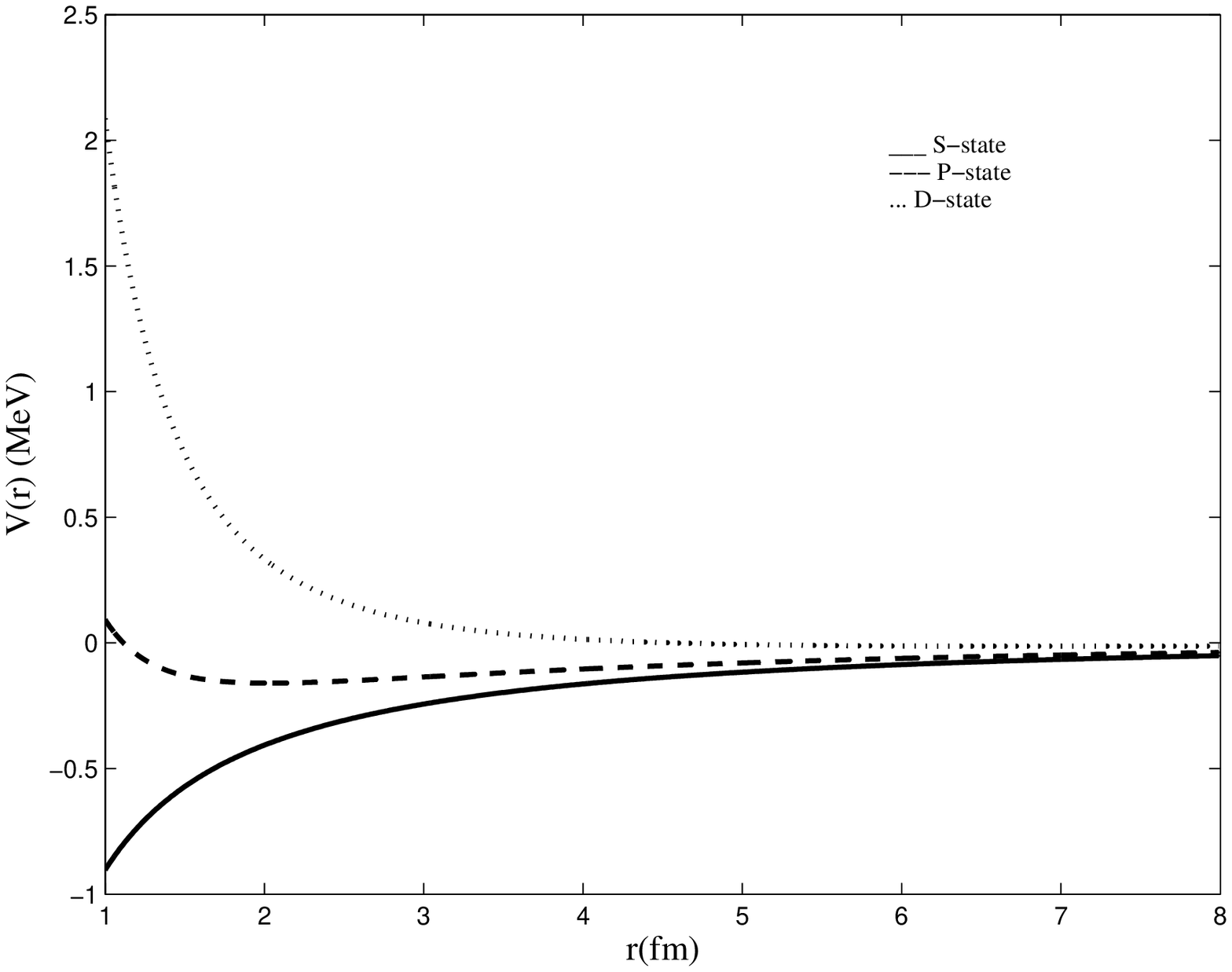,width=17cm,angle=0} \label{Fig1}
\end{figure}

\newpage

\begin{figure}
\epsfig{file=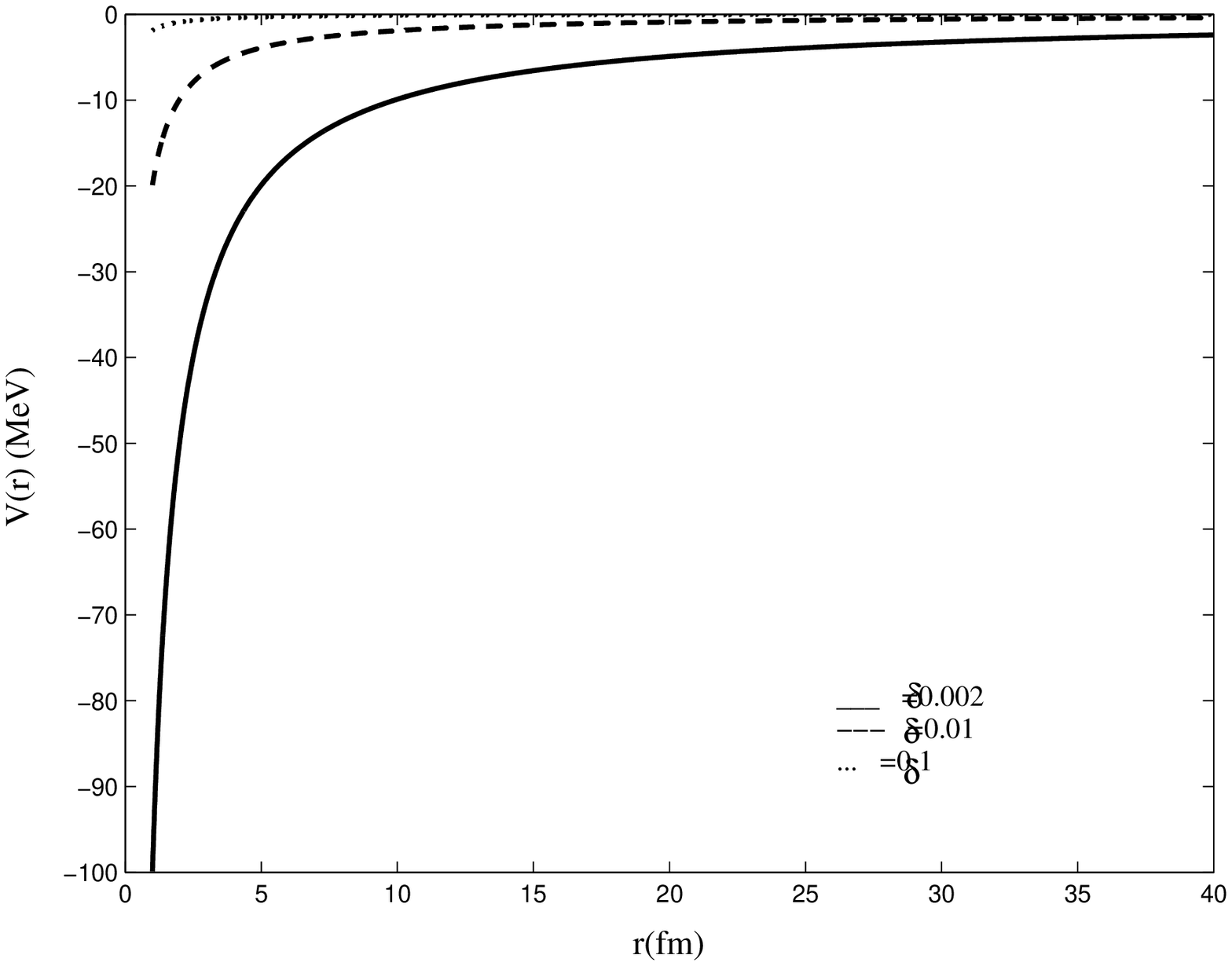,width=17cm,angle=0} \label{Fig2}
\end{figure}

\newpage

\begin{figure}
\epsfig{file=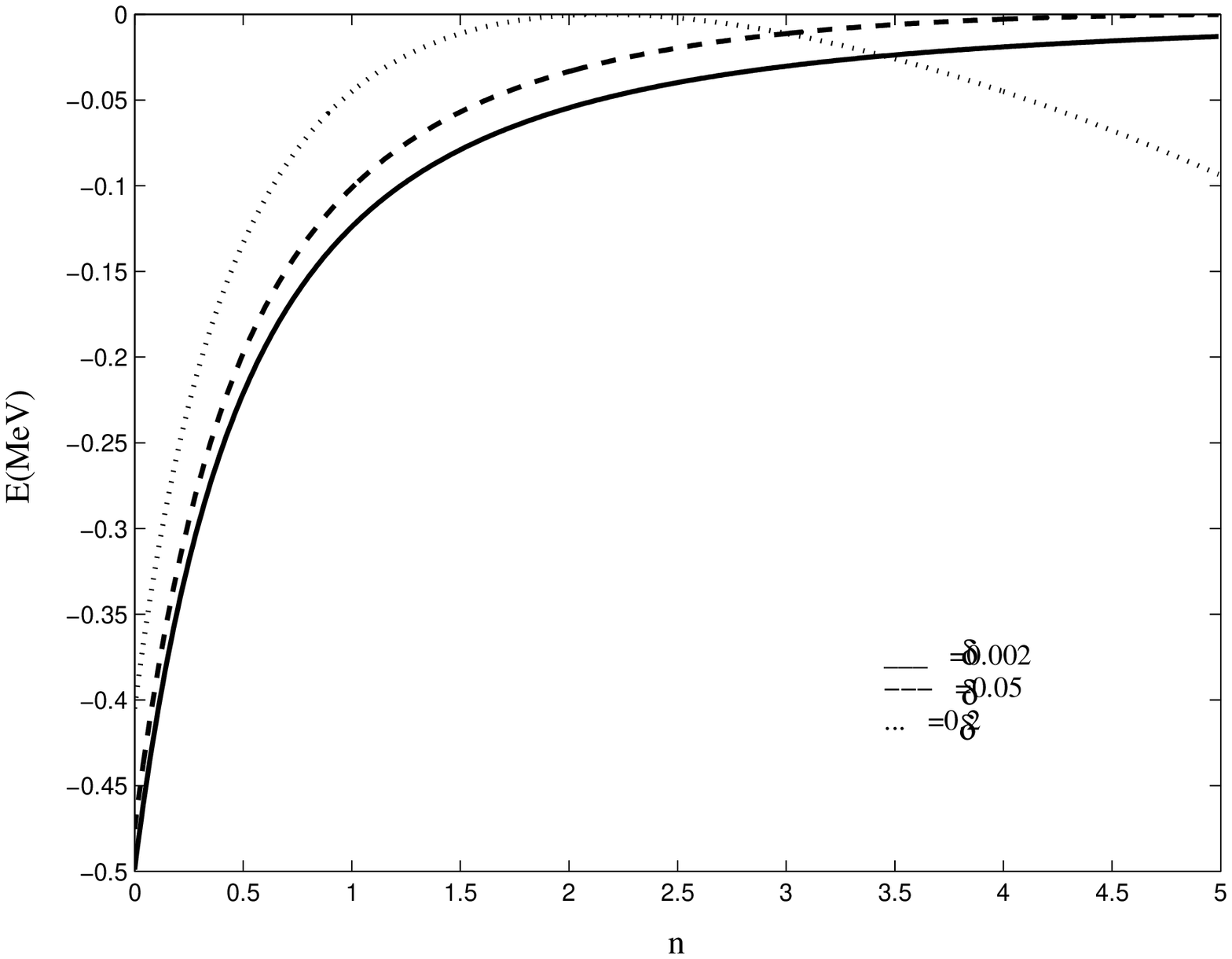,width=17cm,angle=0} \label{Fig3}
\end{figure}

\newpage

\begin{figure}
\epsfig{file=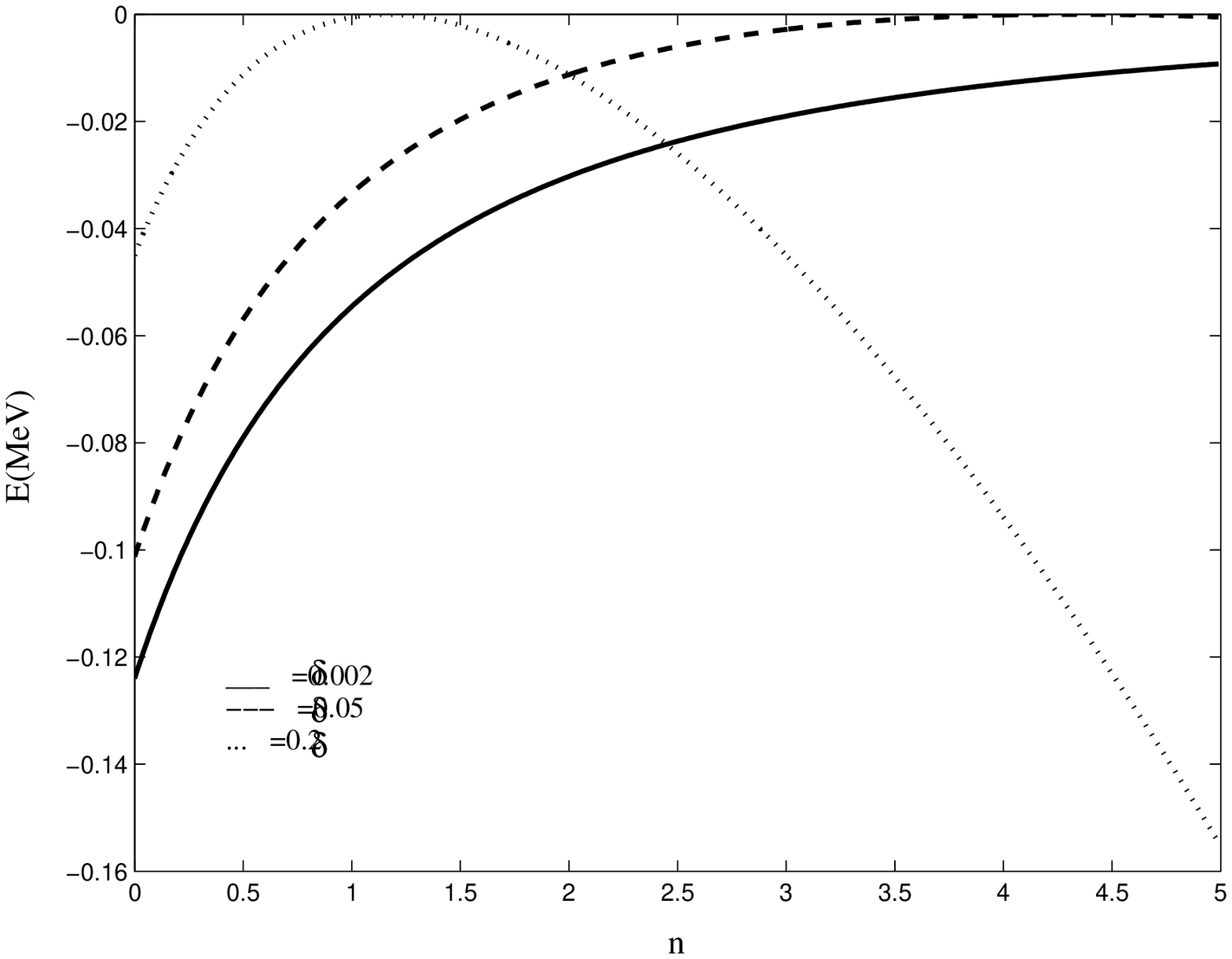,width=17cm,angle=0} \label{Fig4}
\end{figure}
\end{document}